\documentclass[11pt,a4paper]{article}
\usepackage[hyperref]{acl2020}
\usepackage{times}
\usepackage{latexsym}
\usepackage[pdftex]{graphicx}
\usepackage{amssymb}
\usepackage{amsthm}
\usepackage{amssymb}
\usepackage{amsmath}  % math symbols

\usepackage{microtype}

\aclfinalcopy % Uncomment this line for the final submission
\title{DeepEmoNet: Building Machine Learning Models for \\Automatic Emotion Recognition in Human Speeches}

\author{Tai Vu \\
  Department of Computer Science \\
  Stanford University \\
  \texttt{taivu@stanford.edu} 
}

\date{}

\begin{document}
\maketitle

\begin{abstract}
Speech emotion recognition (SER) has been a challenging problem in spoken language processing research, because it is unclear how human emotions are connected to various components of sounds such as pitch, loudness, and energy. This paper aims to tackle this problem using machine learning. Particularly, we built several machine learning models using SVMs, LTSMs, and CNNs to classify emotions in human speeches. In addition, by leveraging transfer learning and data augmentation, we efficiently trained our models to attain decent performances on a relatively small dataset. Our best model was a ResNet34 network, which achieved an accuracy of $66.7\%$ and an F1 score of $0.631$. 
\end{abstract}

\section{Introduction}
In recent decades, the advent of machine learning technologies has accelerated research in spoken language processing. In particular, the applications of neural network architectures like CNNs \citep{lecun1995convolutional}, LSTMs \citep{hochreiter1997long}, and Transformers \citep{vaswani2017attention} have led to major advancements and desirable outcomes in automatic speech recognition and speech synthesis programs.

One interesting area of spoken language research is speech emotion recognition (SER). This problem involves classifying emotions like "happiness" or "anger" based on audio clips of human speeches. This is a highly important task, because enabling computers to understand human emotions can help facilitate communication between humans and machines. However, while there has been significant research in building AI-powered emotion detection systems, closing the gap between AI performance and human performance still proves to be challenging, due to the ambiguity and complexity of human emotions.

Therefore, in this paper, we developed several machine learning models that utilized SVMs, CNNs, and LTSMs for automated emotion classification in human speeches. We also implemented transfer learning and data augmentation techniques during the training process, which allowed our models to achieve good performances with little training data.

\section{Related Works}

Over the last few years, there have been an increasing number of studies on speech emotion recognition (SER). For instance, \citet{schuller2003hidden} leveraged a Hidden Markov Model (HMM) to extract features from speech signals and used them to detect emotions. More recent research has utilized Mel Frequency Cepstral coefficients (MFCCs), which has proven very useful in automatic speech recognition. Specifically, \citet{demircan2018application} extracted MFCCs from the EMO-DB dataset, and then combined them with fuzzy C-means clustering and k-nearest neighbors (kNN) for emotion prediction. 

Meanwhile, together with recent breakthroughs in deep learning, many studies have focused on leveraging the power of neural networks for emotion classification. Particularly, \citet{lim2016speech} applied CNN and LSTM network layers on top of short-time Fourier transform representations of the EMO-DB raw audio data. This approach demonstrated great improvements in predictive accuracy over traditional classification methods. However, most of those deep learning based systems required a large amount of training data in order to achieve high performances. Our project was different, because we trained our machine learning models on a relatively small database. We will demonstrate that incorporating data augmentation and transfer learning can effectively enable our systems to overcome the lack of data, address overfitting issues, and attain decent performances.

In addition, a number of studies have focused on building multimodal systems that harness additional information from videos or texts to improve speech emotion classification. For example, \citet{kim2013deep} combined hand-crafted speech features such as pitch, energy, and mel-frequency filter banks (MFBs) with facial landmark features from videos. On the other hand, \citet{tzirakis2017end} leveraged 1D convolutional layers to encode features from speeches, while using ResNet50 to extract visual information from video frames. The combined features were passed through an LSTM module to perform final prediction. While this multimodal approach led to some improvements in accuracy levels, it is crucial to note that visual and textual information is not always available. Therefore, building audio-only emotion detection systems is highly important for use cases where we only have audio data. This insight motivated us to develop and train our machine learning models to output correct emotion labels solely based on input audio clips without using any visual or textual data.

% (contributions + novelty?)

\section{Approach}

\subsection{Models}

Our machine learning system included an encoder, which was followed by a classifier. The encoder received an audio clip and then produced a vector representation of the input data. Subsequently, this encoding was fed into the classifier, which outputted an emotion label.

\subsubsection*{Model 1: MFCC and SVM}

As a starting point, we implemented the feature extractor using the Mel
Frequency Cepstral Coefficients (MFCC). Afterwards, we took the averages of these MFCC input features across the time dimension and then used them to train a Support Vector Machines (SVM) model \citep{bosertraining} to classify different emotions.

\subsubsection*{Model 2: Log mel spectrograms and LSTM}

Our second model encoded each data point by computing a mel-scaled spectrogram and then converting it to log space. We built an LSTM neural network as our classifier. This network contained $2$ bidirectional LSTM layers, followed by a dropout layer, a linear layer, and a softmax layer.

\subsubsection*{Model 3: Log mel spectrograms and CNN}

\begin{figure}
    \centering
    \includegraphics[width=0.45\textwidth]{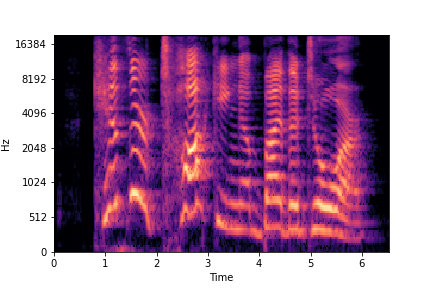}
    \caption{Log mel spectrogram features of an example.}
    \label{fig:logmel}
\end{figure}

In this model, we also extracted log-scaled mel spectrograms for the input speech data. Since these features were similar to 2D image arrays (shown in Figure \ref{fig:logmel}), we then fed them into a CNN classifier in order to obtain emotion labels. Previously, we intended to put raw waveforms directly through the CNN model. However, during our experiments, we found out that training the CNN on log-scaled mel spectrograms was easier and more stable.

We chose ResNet34 \citep{he2016deep} as our CNN architecture. Additionally, we experimented with two different approaches: training a ResNet34 network from scratch and using transfer learning to finetune a ResNet34 model that was pretrained on the ImageNet database \citep{russakovsky2015imagenet}.

% In terms of the encoder, I start with the common MFCC features. After that, I plan to experiment with a combination of CNN or LSTM layers for improved performances. With respect to the classifier, I am implementing the support vector machines (SVMs) algorithm as a starting point. Subsequently, I will implement the classifier using some feedforward neural network layers.

% If I have more time at the end, I would like to explore whether pretrained models like wav2vec can help improve my outcomes. Since, the RAVDESS is quite small, finetuning a neural network pretrained on a big dataset might help boost accuracy levels.

\subsection{Data Augmentation}

As we developed and trained our models on a small speech dataset, data augmentation would be helpful in generating more training data and dealing with overfitting problems.

\subsubsection*{Image-based Data Augmentation}

In particular, since our CNN models were trained on image-like 2D arrays of log-scaled mel spectrograms, we applied several data augmentation methods on these input data, which include rotating by a small degree, zooming in, and changing brightness. Although such image-based augmentation techniques were more common in computer vision tasks and were not directly applied to audio data, we will demonstrate in Section 5.4 that these techniques indeed helped prevent overfitting and improve model performance.

\subsubsection*{Progressive Resizing}

Another augmentation method that we used was progressive resizing \citep{colangelo2021progressive}. Specifically, we first trained the CNN models on smaller versions of the log-scaled mel spectrogram arrays $(128 \times 128)$, and then finetuned the networks on arrays of larger sizes $(256 \times 256)$. This approach not only augmented the training data, but also allowed the models to train much faster.

\subsubsection*{Mixup}

In addition, we harnessed Mixup, a data augmentation technique that generated convex combinations of pairs of training examples and their labels \citep{zhang2018mixup}. Particularly, for two randomly sampled data points $(x_i, y_i)$ and $(x_j, y_j)$, this method constructed a new example of the form:
\begin{equation*}
\begin{aligned}
\tilde x &=& \lambda x_i + (1 - \lambda) x_j \\
\tilde y &=& \lambda y_i + (1 - \lambda) y_j
\end{aligned} 
\end{equation*}

Here, $x_i, x_j$ are input vectors, $y_i, y_j$ are one-hot label encodings, and $\lambda \in [0, 1]$. In this way, Mixup acted as a regularizer that encouraged the linear behaviors of the models, reduced their variance, and enhanced their generalization powers.

\section{Implementation}

I implemented the code for this project in Python using PyTorch \citep{paszke2019pytorch}, FastAI \citep{howard2020fastai}, Scikit-learn \citep{pedregosa2011scikit}, Librosa \citep{mcfee2015librosa}. All the code can be found \href{https://github.com/taivu1998/ML-SER}{here}.

\section{Experiments}

\subsection{Data}

In this project, we used the Ryerson Audio-Visual Database of Emotional Speech and Song (RAVDESS) database \citep{livingstone_steven_r_2018_1188976} and the Surrey Audio-Visual Expressed Emotion (SAVEE) database \citep{jackson2014surrey}. We combined them into a single dataset for training and testing our models.

RAVDESS is an English language database that contains $1440$ utterances. This dataset was made by $24$ actors ($12$ female and $12$ male), who said two sentences "Kids are talking by the door" and "Dogs are sitting by the door" with various emotions. Meanwhile, the SAVEE database consists of $480$ audio clips created by $4$ male actors, and each of them recorded $15$ sentences. There are $8$ different emotion classes, including \texttt{neutral}, \texttt{calm}, \texttt{happy}, \texttt{sad}, \texttt{angry}, \texttt{fearful}, \texttt{disgust}, and \texttt{surprised}.

% \begin{figure}
%     \centering
%     \includegraphics[width=0.5\textwidth]{acl2020-templates/duration.png}
%     \caption{Distribution of recording durations.}
%     \label{fig:duration}
% \end{figure}

\begin{figure}
    \centering
    \includegraphics[width=0.5\textwidth]{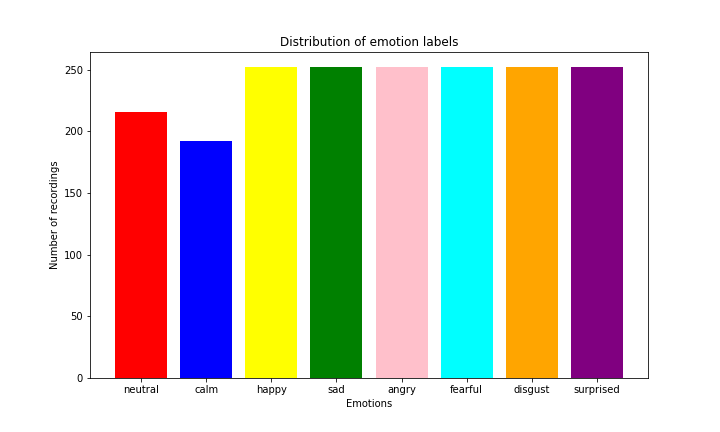}
    \caption{Distribution of emotion labels in the dataset.}
    \label{fig:emotion_dist}
\end{figure}

The duration of each utterance ranges from $3$ to $5$ seconds. The total duration of audio recordings is roughly $2$ hours. In addition, we can see in Figure \ref{fig:emotion_dist} that most of the emotional classes are relatively well balanced. The \texttt{neutral} and \texttt{calm} labels contain slightly fewer audio clips than the other $6$ classes.

% \begin{figure}
%     \centering
%     \includegraphics[width=0.5\textwidth]{acl2020-templates/male_emotion.png}
%     \includegraphics[width=0.5\textwidth]{acl2020-templates/female_emotion.png}
%     \caption{Distribution of emotion labels in each gender group.}
%     \label{fig:label}
% \end{figure}

We split the dataset into $90\%$ for training, $5\%$ for validation, and $5\%$ for testing. 

% In addition, since the size of the dataset is relatively small, I am considering some data augmentation approaches, such as SpecAugment \citep{Park_2019} or Mixup \citep{zhang2018mixup}.

\subsection{Experiment Details}

For the SVM model, we produced $20$ MFCCs for each input audio clip. We chose an RBF kernel for the SVM algorithm.

For the LSTM and CNN models, we generated $128$ mel bands when converting input speeches to mel spectrograms. We trained the LSTM model and the vanilla CNN model (with no pretraining) for $200$ epochs. Meanwhile, for the ResNet34 model that was pretrained on ImageNet, we finetuned its weights for $30$ epochs. We used a batch size of $64$ and a learning rate of $0.001$, with a decay rate of $0.9$. We trained the above neural networks using the Cross Entropy loss and the Adam optimization algorithm \citep{kingma2014adam}.

\subsection{Evaluation Methods}

Since this project tackled a classification problem, we used classification accuracy scores and F1 scores for evaluating model performance.

\subsection{Results}

\begin{table*}
\centering
\begin{tabular}{lll}
\hline
\textbf{Models} & \textbf{Accuracy} & \textbf{F1 Scores}\\
\hline
SVM & 51.7\% & 0.509 \\
LSTM & 52.8\% & 0.497 \\
CNN (trained from scratch) & 45.8\% & 0.426 \\
CNN (transfer learning) & 57.3\% & 0.528 \\
CNN (transfer learning, data augmentation) & \textbf{66.7\%} & \textbf{0.631} \\
\hline
\end{tabular}
\caption{Performance of different models on the validation set.}
\label{results}
\end{table*}

As shown in Table \ref{results}, the SVM algorithm produced an accuracy of $51.7\%$ and an F1 score of $0.509$. This result was better than we expected, because the model only took into account the mean values of the MFCC features across the time dimension. In other words, the SVM algorithm did not get access to useful temporal dependencies amongst the input MFCC features, but still learned to predict emotions with more than $50\%$ accuracy.

After that, the LSTM model performed slightly better than the SVM algorithm, with a higher accuracy of $52.8\%$ and a comparable F1 score of $0.497$. When investigating its training process, we can see that the performance was still quite low because the LSTM network was overfitting to the training data. In particular, the model learned to decrease training losses to a small value (around 0.5), but the validation losses were still high (around 2.9).

A similar pattern occurred for the vanilla CNN model (with no pretraining), as it only produced $45.8\%$ accuracy. In this case, another issue is that because the training set was too small, the Resnet34 network was not able to learn good representations of the speech contents, so it could not generalize well to unseen data.

In fact, when we finetuned the ResNet34 model with pretrained weights from ImageNet, the performance went up significantly ($57.3\%$ in accuracy and $0.528$ in F1 score). Therefore, we can see that the neural network learned useful feature representations of the speech data after being pretrained on a large database like ImageNet. When it was finetuned on our small dataset, the model was able to transfer its prior knowledge about images to reading and extracting information from image-like log-scaled mel spectrogram arrays. The finetuning process then helped the model to adapt to the domain of our dataset even better, which enhanced its performance.

\begin{figure}
    \centering
    \includegraphics[width=0.4\textwidth]{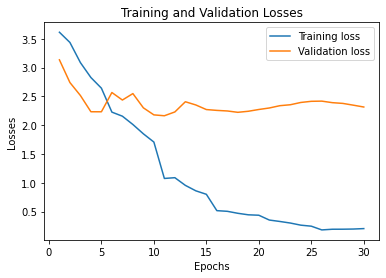}
    \includegraphics[width=0.4\textwidth]{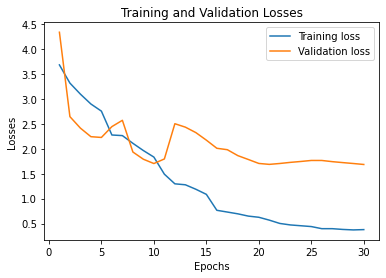}
    \caption{Training and validation losses across 30 epochs for the ResNet34 model without data augmentation (upper) and with data augmentation (lower).}
    \label{fig:cnn_loss}
\end{figure}

Finally, the ResNet34 model with both transfer learning and data augmentation achieved the best performance, with an accuracy of $66.7\%$ and an F1 score of $0.631$. This illustrates the effectiveness of data augmentation techniques in boosting our model performance. Indeed, as we can see in the upper plot of Figure \ref{fig:cnn_loss}, the ResNet34 network without data augmentation was still overfitting, with low training loss values and high validation loss values. This means that the gap between the training losses and the validation losses was still very large. However, this problem was alleviated with the support of data augmentation, as shown in the lower plot of Figure \ref{fig:cnn_loss}. Both the training losses and the validation losses decreased gradually, and the gap between them was significantly narrowed.

% - LSTM model + CNN model no pretrained: not too good, overfitting, not sufficient data for learning good representation of speech content

% - Transfer learning helps: pretrained model learn how to extract features from images, then finetuning help it adapt

\begin{figure}
    \centering
    \includegraphics[width=0.35\textwidth]{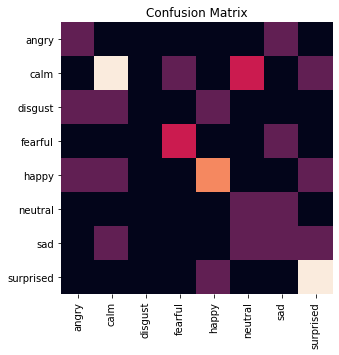}
    \caption{Confusion matrix for the best CNN model.}
    \label{fig:confusion_matrix}
\end{figure}

\begin{figure}
    \centering
    \includegraphics[width=0.45\textwidth]{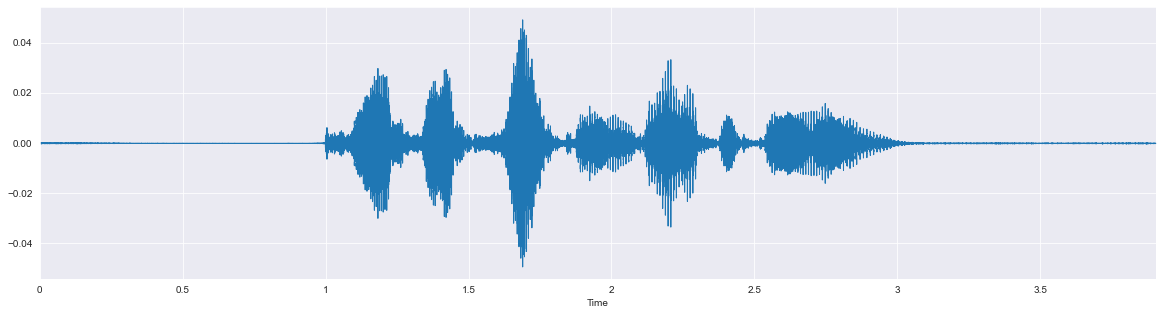}
    \includegraphics[width=0.45\textwidth]{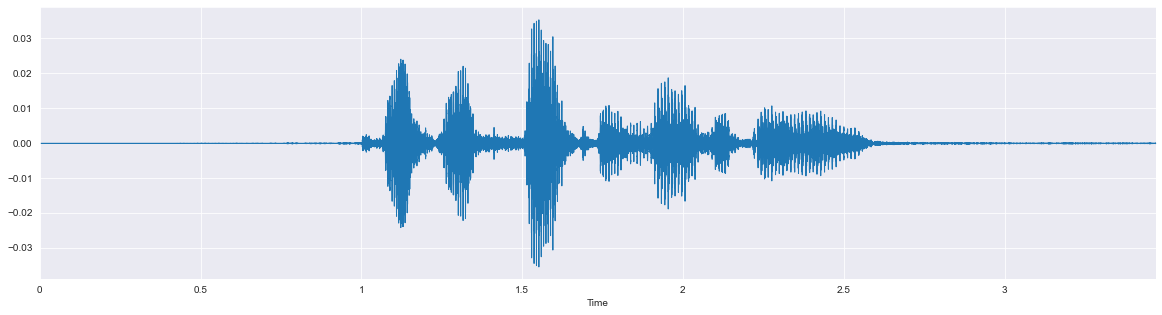}
    \caption{The waveforms of a neutral utterance (upper) and a calm utterance (lower) from the same actor.}
    \label{fig:neutral_calm}
\end{figure}

Meanwhile, because the accuracy of our final model was less than $70\%$, there is still a lot of room for improvement. One of the main challenges faced by our models was that RAVDESS and SAVEE were two simulated datasets, which consisted of several actors repeating the same sentences with various emotions. Hence, the speech contents in these datasets were not diverse enough for our machine learning programs to learn proper representations of input audio data and detect correlations between human speeches and emotions. In addition, we can observe in Figure \ref{fig:confusion_matrix} that the ResNet34 model performed well on certain positive classes like \texttt{surprised}, \texttt{happy}, and \texttt{calm}, while produced lower accuracy on some other negative classes like \texttt{disgust} and \texttt{angry}. Furthermore, there was some confusion between certain pairs of emotion labels, such as \texttt{neutral} and \texttt{calm}. This issue is understandable, because the audio clips from these two classes in our dataset often sound similar. Two examples from those two classes are shown in Figure \ref{fig:neutral_calm}.

% challenging (do well on happy and suprised, other like neutral or sad is not as good, Confusion between classes - ambiguous - examples of neutral and calm, stimulated dataset - similar speech contents)

% Although RAVDESS is a relatively popular benchmark dataset, most research have focused on evaluations of multimodal approaches on this database. One baseline result for an audio-only speech emotion recognition system is provided in \citet{beard-etal-2018-multi}. Specifically, their model achieved a test accuracy of $41.25\%$.

% I am currently building a MFCC feature extractor and an SVM classifier, and will obtain some results soom for comparison.

% \subsection{Analysis}

\section{Conclusion}

Overall, in this study, we developed a number of machine learning models, including SVMs, LSTMs, and CNNs, for inferring emotions from human speeches. Our models were trained and evaluated on small dataset created from the RAVDESS and SAVEE databases. Our best model was a ResNet34 neural network, which achieved an accuracy of $66.7\%$ and an F1 score of $0.631$. This is a promising result, given the small size of our training set. With more training data, the model will definitely be able to learn better and recognize emotion classes with higher accuracy levels. In addition, we demonstrated the benefits of transfer learning and data augmentation in boosting model performance. Particularly, transfer learning allowed the model to overcome the lack of audio data and learn good feature representations of speech contents, while data augmentation helped create more training examples, prevent overfitting issues, and enhance the robustness and generalization of the model.

The next step would be performing more hyperparameter tuning in order to improve our current models. Additionally, we are interested in experimenting with a combination of CNN or LSTM layers for better performances. Furthermore, given the great advantages of data augmentation, we want to implement several audio-based data augmentation techniques such as pitch shift, change in loudness, change in speed, and SpecAugment \citep{Park_2019}, as they might be able to further reduce overfitting and generalization errors in our training pipeline. Finally, because transfer learning is also beneficial, we would like to finetune some pretrained speech models such as wav2vec \citep{schneider2019wav2vec} and SpeechBERT \citep{chuang2019speechbert}, and see how they perform in the speech emotion recognition task.

% \section{Contribution}

% Because this is a solo project, I (Tai Vu) implemented the entire code for the project, including data preprocessing, data pipeline, training pipeline, machine learning models, and evaluation.

\newpage
\bibliography{anthology,acl2020}
\bibliographystyle{acl_natbib}

% \begin{thebibliography}{10}
% \bibitem{iemocap}
% Busso, Carlos, et al. "IEMOCAP: Interactive emotional dyadic motion capture database." Language resources and evaluation 42.4 (2008): 335-359.
% % https://sail.usc.edu/iemocap/

% \bibitem{ravdess}
% Livingstone, Steven R., and Frank A. Russo. "The Ryerson Audio-Visual Database of Emotional Speech and Song (RAVDESS): A dynamic, multimodal set of facial and vocal expressions in North American English." PloS one 13.5 (2018): e0196391.
% % https://zenodo.org/record/1188976\#.YC3q1y2z3yz

% \bibitem{savee}
% Jackson, Philip, and S. Haq. "Surrey audio-visual expressed emotion (savee) database." University of Surrey: Guildford, UK (2014).
% % http://kahlan.eps.surrey.ac.uk/savee/

% \end{thebibliography}

\end{document}